\documentclass{emulateapj-rtx4}
\usepackage{apjfonts}
\usepackage{graphicx}
\usepackage{amssymb}
\usepackage{amsbsy}
\usepackage{amsmath}
\usepackage{mathrsfs}
\usepackage{color}

\def\Lb{\mathbf{L}}
\def\elb{\boldsymbol{\ell}}
\def\Jhb{\mathbf{J}_{\rm h}}
\def\Jdb{\mathbf{J}_{\rm d}}
\def\rd{{\rm d}}

\slugcomment{The Astrophysical Journal, 2013, 764, 16}
\shortauthors{LI ET AL.}
\shorttitle{EVOLUTION OF WARPED ACCRETION DISKS IN AGNs. I. }
\begin{document}

\title{Evolution of Warped Accretion Disks in Active Galactic Nuclei. I. Roles of Feeding at the Outer Boundaries}
\author{
Yan-Rong Li\altaffilmark{1}, Jian-Min Wang\altaffilmark{1,2} , Cheng Cheng\altaffilmark{1} and 
Jie Qiu\altaffilmark{1}
}
\altaffiltext{1}
{
Key Laboratory for Particle Astrophysics, Institute of High 
Energy Physics, Chinese Academy of Sciences, 19B Yuquan Road, 
Beijing 100049, China; liyanrong@mail.ihep.ac.cn
}

\altaffiltext{2}
{
National Astronomical Observatories of China, Chinese 
Academy of Sciences, 20A Datun Road, Beijing 100020, China
}

\begin{abstract}
We investigate the alignment processes of spinning black holes and their surrounding
warped accretion disks in a frame of two different types
of feeding at the outer boundaries. We consider (1) {\em fixed flows} in 
which gas is continually fed with a preferred angular momentum, and 
(2) {\em free flows} in which there is no gas supply and the disks diffuse freely 
at their outer edges. As expected, we find that for the cases of fixed flows the
black hole disk systems always end up aligning on timescales of several $10^6$ 
yr, irrespective of the initial inclinations. If the initial inclination angles are larger
than $\pi/2$, the black hole accretion transits from retrograde to prograde fashion, 
and the accreted mass onto the black holes during these two phases is comparable.
On the other hand, for the cases of free flows, both alignments and anti-alignments
can occur, depending on the initial inclinations and the ratios of the angular momentum 
of the disks to that of the black holes. In such cases, the disks will be consumed within
timescales of $10^6$ yr by black holes accreting at the Eddington limit. 
We propose that there is a close connection between the black hole spin and the lifetime for which
the feeding persists, which determines the observable episodic lifetimes of active 
galactic nuclei. We conclude that careful inclusion of the disk feeding at the 
outer boundaries is crucial for modeling the evolution of the black hole spin.
\end{abstract}


\section{Introduction}
The current hierarchical framework of galaxy formation and evolution predicts that, 
in addition to secular evolution, repeat galaxy mergers trigger activities of 
supermassive black holes (SMBHs) located at the galactic center and hence their 
growth (e.g., \citealt{Benson2010}). As such, fuel feedings channeled onto SMBHs  
most likely proceed at episodic and random phases (e.g., \citealt{Martini2004, King2006, 
Wang2006, Wang2008, Wang2009, Li2010}). This picture is further reinforced by 
various lines of observations, such as misaligned active galactic nuclei (AGNs) with 
respect to their host galaxies (\citealt{Kiney2000, Gallimore2006, MunozMarin2007, Shen2010}),
and as well as by numerical simulations (e.g., \citealt{Hopkins2012} and references therein).
Previous studies on the cosmological evolution of the radiative efficiency ($\eta$) of accretion
according to the \citeauthor{Soltan1982}'s (1982) argument and formulated by the
$\eta-$equation, found that $\eta$ decreases with 
cosmic time since redshift $z\sim2$, strongly implying that random accretion 
takes place upon the SMBHs  (\citealt{Wang2009, Li2011,
Li2012}; see also the discussion of \citealt{Zhang2012}).  
A similar inference has recently been highlighted  
from the semi-analytical modeling on SMBH demography by \cite{Volonteri2012}.
In this context, warped accretion disks are ubiquitous in AGNs.

A misaligned accretion disk around the central spinning black hole undergoes 
Lense-Thirring precession arising from the gravitomagnetic effects, of which 
the rate falls off rapidly with the radius to the hole (e.g., \citealt{Hartle2003}). 
The presence of viscosity combined with such differential precession
will induce the inner portion of the disk to align or anti-align its orbital angular 
momentum with the spin axis of the hole out to a transition radius, beyond 
which the disk retains its initial inclination (Bardeen-Petterson effect; 
\citealt{Bardeen1975}).  In the meantime, associated with disk precession, 
the black hole suffers an equal and opposite gravitomagnetic torque that causes 
it to precess as well. Under mutual interaction and warp propagation in the disk, 
this composite system tends to restore full axisymmetry with the entire
disk ultimately aligned or anti-aligned with the hole (\citealt{King2005}).
The characteristic timescale and the final state of the system (alignment or anti-alignment) depend on the 
black hole mass and spin and the properties of the accretion disk.

Based on the angular momentum {\em conservation}, \cite{King2005} demonstrated 
that the disk will end up anti-aligned with the hole if the angular momentum of 
the hole ($J_{\rm h}$) dominates over that of the disk ($J_{\rm d}$). 
This refreshes previous analytical studies on the grounds that the disk has an 
infinite extension and is continuously fed (\citealt{Scheuer1996, Martin2007}), and 
has been incorporated into the black hole spin modeling in the context of hierarchical 
cosmology (\citealt{King2008, Lagos2009, Fanidakis2011, Barausse2012}).
However, while $J_{\rm h}$ is accurately defined, the meaning of $J_{\rm d}$ is  
not as well defined in \cite{King2005}, leading to its interpretation varying among 
subsequent studies.

It is commonplace that, in the outer region, the accretion disk likely becomes 
self-gravitating, fragments into clumps, and probably manifests itself as a 
star-forming region (e.g., \citealt{Paczynski1978, Shlosman1987, Collin1999, 
Goodman2003, Wang2010}).  The feedback from the stellar radiation or the 
resultant supernova explosions embedded in the disk acts to excite turbulence 
that continues to deprive gas of angular momentum (\citealt{Wada2002, Kawakatu2008, 
Kumar2010, Wang2010}). In this case, it is expected that gas will be continually 
fueled into the accretion disk from the outer region (e.g., circumnuclear disk or torus)
for the duration of the AGN's lifetime. The practical implications are twofold: an outer edge 
of the disk limited by its own self-gravity exists so that the angular momentum of the disk
can be properly defined, and there is a gas supply with specified angular momentum 
at this outer edge.

With the above inference, a comprehensive study of the alignments/anti-alignments 
between black holes and their warped accretion disks is essential, in particular 
considering the increased focus on modeling the black hole spin across cosmic time.
In this work, we revisit this issue by numerically solving the evolution equation 
set of black holes and accretion disks in a practical sense. The paper is organized as follows: 
Section~2 describes the evolution equation set for a black hole and its warped accretion 
disk. Section~3 surveys the properties of warped accretion disks from the theoretical 
point of view. Section 4 shows the numerical scheme and sets up the boundary conditions 
and initial conditions for solving the evolution equation set. The results are then 
presented in Section~5. The implications of our results on SMBH spin evolution
and the uncertainties in our calculations are discussed in Section~6. Conclusions are
summarized in Section~7.

\begin{figure*}[t!]
\centering
\includegraphics[height=2.4cm]{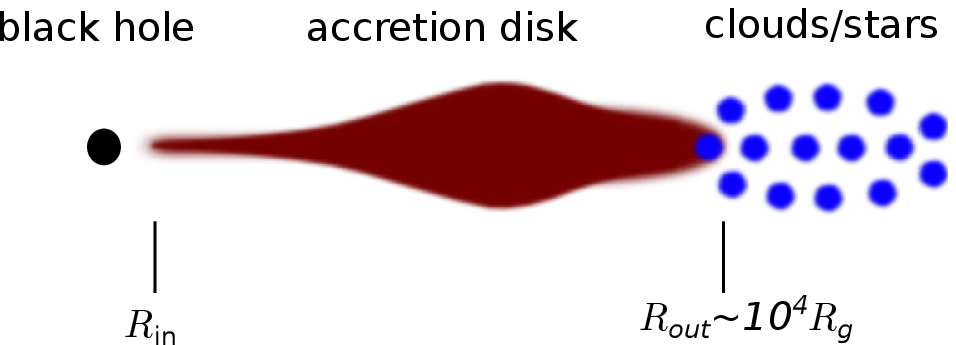}\hspace*{2.5cm}
\includegraphics[height=2.4cm]{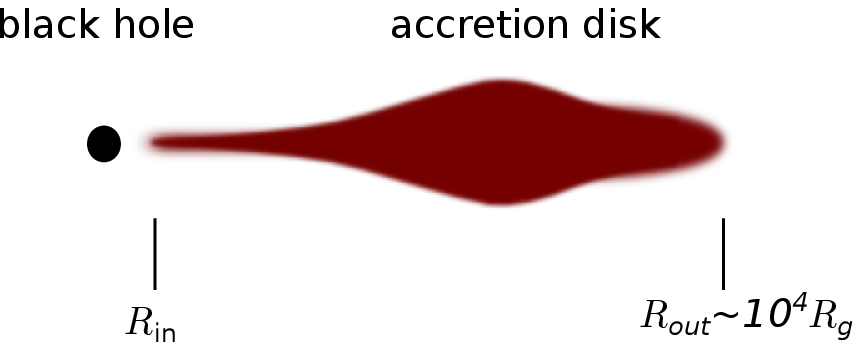}
\caption{Schematic illustration of two classes of accretion models.
Left: {\em fixed flows} in which gas is continually fed with a preferred angular momentum 
through the outer region composite of cloud and stars.
Right: {\em free flows} in which there is no gas supply and the disk diffuses freely at its outer edge,
leading to an outflow of angular momentum.
The central black hole spin is oriented with an inclination to the accretion disk.
The outer edge of the accretion disk is limited by self-gravity of the disk and the inner
edge is determined by the marginal stable orbit.}
\label{fig_sch}
\end{figure*}

\section{Basic Evolution Equations}

The dynamics of warped disks have been studied extensively through theoretical analysis 
(e.g., \citealt{Papaloizou1983, Pringle1992, Papaloizou1995, Ogilvie1999, Ogilvie2000, Lubow2002})
and numerical simulations (e.g. \citealt{Larwood1996, Nelson1999, Nelson2000, Dotti2010, Lodato2010, 
Nixon2012, Nixon_etal2012, McKinney2013}).
Adopting the standard $\alpha-$prescription of disk viscosity, warp propagation in the disk is characterized
into two regimes depending on the relative importance of pressure forces and viscous forces.
For a nearly inviscid or sufficiently thick disk such that the dimensionless viscosity coefficient $\alpha$ 
is smaller than the disk aspect ratio $h=H/R$, the warp propagates as bending waves at
a velocity related to the speed of sound (\citealt{Papaloizou1995});  
whereas in the regime $\alpha>h$, the warp evolves in a diffusive fashion.
Here $H$ is the semithickness of the disk. 

We focus our attention on a thin and viscous accretion disk with a Keplerian rotation ($\alpha>h$),
the most likely case for accretion disks in AGNs (\citealt{Pringle1999}).
The evolution of such warped disks can be completely described by 
the angular momentum density of the disk $\Lb(R, t)$ as (\citealt{Ogilvie1999})
\begin{eqnarray}\nonumber
\frac{\partial \Lb}{\partial t}+\frac{1}{R}\frac{\partial (V_R R\Lb)}{\partial R}
&=&-\frac{1}{R}\frac{\partial}{\partial R}\left(\frac{3}{2}\nu_1\Lb\right)
       +\frac{1}{R}\frac{\partial}{\partial R}\left(\frac{1}{2}\nu_2R L \frac{\partial \elb}{\partial R}\right)\\
&&  +\frac{1}{R}\frac{\partial}{\partial R}\left(\nu_3 R \Lb\times\frac{\partial \elb}{\partial R}\right)
       +\mathbf{T}_{\rm LT}, 
\label{equ_disk}
\end{eqnarray}
where $L=\Sigma R^2\Omega$, $\elb=\Lb/L$ denotes the direction vector, 
$\Sigma$ is the mass surface density, $\Omega$ is the angular velocity, $\mathbf{T}_{\rm LT}$
is the gravitomagnetic torque responsible for the Lense-Thirring precession, and 
$V_R$ is the radial velocity given by
\begin{equation}
V_R=-\frac{3}{L}\frac{\partial (\nu_1 L)}{\partial R} -\nu_2 R \left|\frac{\partial \elb}{\partial R}\right|^2.
\label{equ_vr}
\end{equation}
Here the internal viscous torque 
of the disk fluid has been distinguished into three kinds associated with three effective viscosities,
respectively (\citealt{Bardeen1975, Ogilvie1999}). The first viscosity, $\nu_1$, governs the usual azimuthal 
shear due to the differential rotation;
the second viscosity, $\nu_2$, governs the vertical shear that acts to diffuse the warp throughout
the disk; and the third viscosity, $\nu_3$, governs a torque component that tends to create the disk ring
process if it is misaligned with its neighbors. The later two viscosities disappear in a  flat disk.
It is worth pointing out that Equation (\ref{equ_disk}) remains valid for large amplitude warp
(\citealt{Ogilvie1999}).

By analyzing the internal non-linear dynamics of warped accretion disks, \cite{Ogilvie1999}
formulated the effective viscosities and showed that they generally depend on the amplitude of the warp.
For the sake of simplicity, we make use of the first-order approximations as follows
\begin{equation}
\nu_1=\alpha  h^2R^2\Omega,
\label{equ_vis}
\end{equation}
$\nu_2=\nu_1/2\alpha^2$ and $\nu_3=3\nu_1/8\alpha$. As a result,  the warp propagates in the disk on 
a timescale of an order of 
\begin{equation}
t_{\rm w}=\frac{R^2}{\nu_2}\sim 2\alpha^2\frac{R^2}{\nu_1}=2\alpha^2 t_{\nu},
\label{equ_tw}
\end{equation}
where $t_\nu$ is the usual viscous timescale.

In the weak field limit, the gravitomagnetic torque due to frame dragging has a form of 
(\citealt{Bardeen1975})
\begin{equation}
\mathbf{T}_{\rm LT}=\frac{2G}{c^2}\frac{\Jhb\times\Lb}{R^3},
\label{equ_torque}
\end{equation}
where $\Jhb$ is the angular momentum of the hole with its modulus $J_{\rm h}=aG M_\bullet^2/c$,
$a$ is the spin parameter, $M_\bullet$ is the mass of the black hole, $G$ is the gravity constant, 
and $c$ is the speed of light. Due to the mass accretion and the gravitomagnetic torque exerted by 
the warped accretion disk, the angular momentum of the black hole evolves with time as
\begin{equation}
\frac{\rd\Jhb}{\rd t}=\dot M_{\rm in}\mathbf{j}_{\rm in} -2\pi \int \mathbf{T}_{\rm LT} R \rd R, 
\label{equ_bh}
\end{equation}
where $\dot M_{\rm in}$ is the mass accretion rate through the inner edge of the disk $R_{\rm in}$,
which is determined by the marginal stable orbit, and $\mathbf{j}_{\rm in}$ is the specific 
angular momentum carried by matter at the inner edge.

To study the roles of disk feeding,  we consider two classes of accretion models to mimic 
the realistic accretion disks, as schematically shown in Figure~\ref{fig_sch}. In the left panel, 
an outer clumpy region composite of clouds and/or stars continues to channel gas to the 
accretion disk from the outer environment (e.g., circumnuclear disk or torus). 
As stated above, the feedback 
from stellar radiation and supernova explosions are the potential mechanism responsible for 
exciting turbulence and viscosity, acting to transport the angular momentum of accreting matter
(\citealt{Wada2002, Kumar2010, Wang2010}). We denote this case as ``fixed flows'' because the 
angular momentum of gas supply at the outer edge has a preferred direction. By contrast,
the right panel of Figure \ref{fig_sch} shows the other class of outer boundary conditions,
in which there is no gas supply and the disk diffuses 
freely at its outer edge. We therefore denote this case as ``free flows''.
Before numerically solving the above equations, we endeavor to explore the properties
of warped accretion disks from a theoretical point of view in the next section.

\section{Warped Accretion Disks}
We start by specifying the accretion disk model used in this paper generally following 
\cite{Collin-Souffrin1990}. To simplify the theoretical analysis and numerical calculations below, 
we parameterize accretion disks with a Keplerian rotation of $\Omega=\sqrt{GM_\bullet/R^3}$,
and a constant aspect ratio of $h=H/R$. The mass surface density of the disk obeys
$\Sigma=\Sigma_0R^{-p}$, where $\Sigma_0$ is a coefficient determined by setting
the total disk mass  at a fraction of the black hole mass $M_{\rm d}=f_m M_\bullet$, and 
$p=1/2$ is an index to enforce the mass accretion rate of $\dot M\approx 3\pi\nu_1\Sigma$ to a constant
at a large radius as for a flat disk. The outer edge of the disk $R_{\rm out}$ is located where 
self-gravitating becomes important, whose value depends on the mass accretion rate and black hole mass
(e.g., \citealt{Goodman2003, King2008}).
In what follows, for illustration purposes,  we adopt fiducial values of $\alpha=0.1$,
$h=10^{-2}$, $f_m=10^{-2}$, $R_{\rm out}=10^4R_{\rm g}$, $M_\bullet=10^8M_\odot$, and 
$a=0.998$, where $R_{\rm g}=GM_\bullet/c^2$ is the gravitational radius.

\subsection{Transporting Angular Momentum }
In planar accretion disks, matter gradually falls inward the central black hole, whereas nearly all of the angular
momentum is transported outward due to viscosities (\citealt{Frank1992}). This remains valid for 
warped accretion disks (\citealt{Scheuer1996}). 
The difference is that gravitomagnetic interactions of the hole and the disk
induce extra transfers of angular momentum between them. 
This transfer rate depends on the Lense-Thirring precessing rate and warp propagation velocity.
There exists a transition radius where  the Lense-Thirring precessing timescale is comparable with 
the warp propagation timescale, which is usually denoted by a ``warp radius''. 

Within the warp radius, the Lense-Thirring precessions are dominant. Viscosities damp out the
differential precessions rapidly, giving rise to an inner flat disk. 
It is easy to verify:
if initially the angle of the disk and the hole is smaller than $\pi/2$, the inner flat disk is aligned with the hole.
By contrast,  if the angle is larger than $\pi/2$, the inner flat disk is anti-aligned.
In the region around the warp radius, the incoming matter is inclined toward the inner flat disk, and the
associated angular momentum is transferred to the hole through gravitomagnetic interaction.
Further beyond the warp radius, warps propagate outward efficiently and the misaligned angular 
momentum is transported outward as well.  

Since the gravitomagnetic torque integrated over the whole disk in Equation (\ref{equ_bh}) 
is always orthogonal to $\Jhb$, it can be expressed in the form of (\citealt{King2005})
\begin{equation}
\mathbf{T}=  -2\pi \int \mathbf{T}_{\rm LT} R \rd R=
-K_1{\hat {\mathbf J}}_{\rm h}\times {\hat {\mathbf J}}_{\rm d}
-K_2{\hat{\mathbf J}}_{\rm h}\times({\hat {\mathbf J}}_{\rm h}\times{\hat {\mathbf J}}_{\rm d}),
\label{equ_torqueal}
\end{equation}
where a hat symbol on top of a vector stands for the corresponding direction vector, 
$K_1$ and $K_2$ are time-dependent coefficients, and
\begin{equation}
\Jdb=2\pi\int\Lb R\rd R,  
\end{equation}
is the total angular momentum of the disk. Here, the first term corresponds to the precessions and
the second term corresponds to changes in the angle between the hole and the disk.
Below we will numerically demonstrate that  $K_1$ and $K_2$ are generally positive 
(see also \citealt{Scheuer1996, King2005, Martin2007}). 

\subsection{Magnitude Estimates}
In in previous section, we demonstrate that the characteristic extension of warps is determined by 
the timescales for warp propagation given
by Equation (\ref{equ_tw}) and for the local Lense-Thirring precessing 
$t_{\rm LT}\approx |\Lb|/|\mathbf{T}_{\rm LT}|=c^2R^3/2GJ_{\rm h}$ 
(\citealt{Scheuer1996}). 
By equating these two timescales,  one obtains the warp radius
$R_{\rm w}=2GJ_{\rm h}/\nu_2c^2$. Using the formulation of $\nu_2$
and the fiducial values of parameters, $R_{\rm w}$
is of the order of 
\begin{eqnarray}
\frac{R_{\rm w}}{R_{\rm g}}=\left(\frac{4\alpha a}{h^{2}}\right)^{2/3}
=250a^{2/3}\left(\frac{\alpha}{0.1}\right)^{2/3}\left(\frac{h}{10^{-2}}\right)^{-4/3}.
\label{equ_rwarp}
\end{eqnarray}

Considering that the Lense-Thirring torque in Equation (\ref{equ_torque}) falls off rapidly with
the radius in proportion to $R^{-3}$, the major contribution it has on the hole comes from
the region of the disk around $R_{\rm w}$. As a result, the torque that the disk exerts on the hole
approximates 
\begin{equation}
 T_{\rm LT}\sim\frac{4\pi G}{c^2}\frac{J_{\rm h} L(R_{\rm w})}{R_{\rm w}}
 \sim \frac{\nu_2}{\nu_1}\dot M \sqrt{GM_\bullet R_{\rm w}},
\end{equation}
where the mass accretion rate $\dot M\sim3\pi\nu_1\Sigma$ is used for simplicity. This results in a timescale for the alignment
between the hole and the whole disk (see also \citealt{Perego2009})
\begin{eqnarray}
t_{\rm al}&=&\frac{J_{\rm h}}{T_{\rm LT}}
\sim a\frac{\nu_1}{\nu_2}\frac{M_\bullet}{\dot M}\sqrt{\frac{R_{\rm g}}{R_{\rm w}}}\nonumber\\
&\sim&10^{-3}a^{2/3}t_{\rm gr}\left(\frac{\alpha}{0.1}\right)^{5/3}\left(\frac{h}{10^{-2}}\right)^{2/3},
\label{equ_tal}
\end{eqnarray}
where $t_{\rm gr}=M_\bullet/\dot M$ is the growth timescale for a black hole accreting at a rate of $\dot M$. 
Again, with the help of the approximation $\dot M\sim3\pi\nu_1\Sigma$, one can estimate 
the mass accretion rate as
\begin{eqnarray}
\dot M&\sim&\frac{c}{R_g}\frac{\alpha f_m M_\bullet h^2}{(R_{\rm out}/R_g)^{3/2}}
\sim10^{-8}M_\bullet{\rm~yr^{-1}}\nonumber\\
&&\times \left(\frac{\alpha}{0.1}\right)\left(\frac{h}{10^{-2}}\right)^2
 \left(\frac{f_m}{10^{-2}}\right)\left(\frac{R_{\rm out}}{10^{4}R_{\rm g}}\right)^{-3/2},
 \label{equ_mdot}
\end{eqnarray}
and the growth timescale $t_{\rm gr}\sim10^8$yr. We note that this estimated accretion rate
is generally equal to the Eddington limit.

If there is no gas supply to the accretion disk,
the disk will be consumed  on a timescale of an order of 
\begin{eqnarray}
t_{\rm d}&=&\frac{M_{\rm d}}{\dot M}=\frac{f_m M_\bullet}{\dot M}\sim10^6 {\rm yr}
 \left(\frac{\alpha}{0.1}\right)^{-1}\left(\frac{h}{10^{-2}}\right)^{-2}\left(\frac{R_{\rm out}}{10^{4}R_{\rm g}}\right)^{3/2},
\label{equ_td}
\end{eqnarray}
which is independent of $f_m$ and $M_\bullet$. 
As can be seen from Equation~(\ref{equ_tal}), once the disk is gradually depleted (without gas replenishment),
the resulting decreases of surface density and the mass accretion rate will significantly prolong
the alignment processes. 
The timescale for warp propagation throughout the disk in Equation (\ref{equ_tw}) is
\begin{eqnarray}
t_{\rm w}&=&\frac{R_{\rm out}^2}{\nu_2}=2\alpha h^{-2}\Omega^{-1}
=3\times10^4{\rm yr}\nonumber\\
&&\times\left(\frac{\alpha}{0.1}\right)\left(\frac{h}{10^{-2}}\right)^2
\left(\frac{M_\bullet}{10^8M_\odot}\right)\left(\frac{R_{\rm out}}{10^4R_{\rm g}}\right)^{3/2}.
\label{equ_tw2}
\end{eqnarray}
The total angular momentum of the disk is
\begin{equation}
J_{\rm d}=|\Jdb|\approx\frac{3}{4} f_mM_\bullet \sqrt{GM_\bullet R_{\rm out}}. 
\end{equation}
This gives a ratio of $J_{\rm d}$ over $J_{\rm h}$:
\begin{equation}
\frac{J_{\rm d}}{J_{\rm h}}\approx\frac{3}{4} \frac{f_m}{a}\sqrt{\frac{R_{\rm out}}{R_{\rm g}}}
=\frac{3}{4}a^{-1}\left(\frac{f_m}{10^{-2}}\right)\left(\frac{R_{\rm out}}{10^4R_{\rm g}}\right)^{1/2}.
\label{equ_ratio}
\end{equation}
\begin{figure*}[t!]
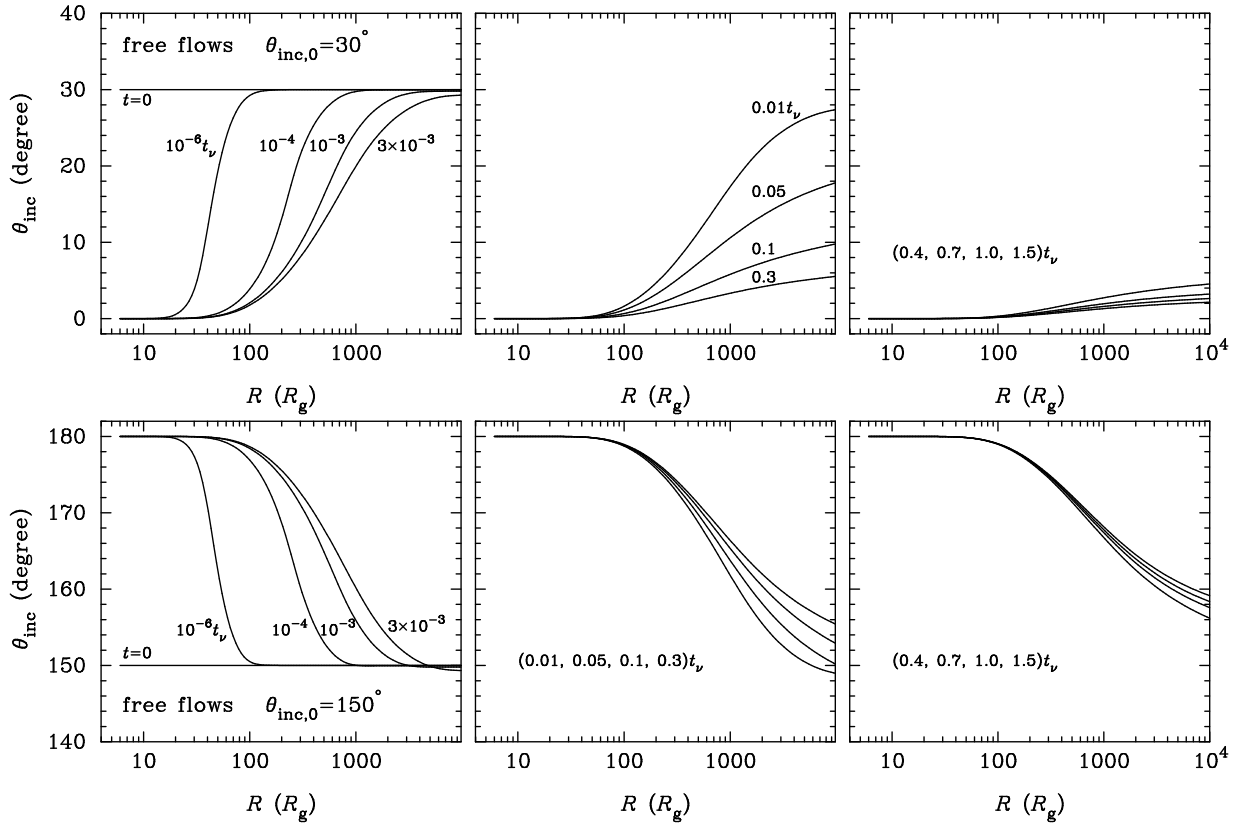

 \centering
 \includegraphics[angle=-90.0, width=0.9\textwidth]{profile11.ps}
 \includegraphics[angle=-90.0, width=0.9\textwidth]{profile21.ps}
 \caption{Shape of accretion disks with initial inclinations of $\theta_{\rm inc,0}=30^\circ$ (top) 
  and $150^\circ$ (bottom)  for free flows at various times.
  Here $t_{\nu}=2\times10^6$yr is the viscous timescale at $R_{\rm out}$ (see Equation (\ref{equ_tw})). }
  \label{fig_profile1}
\end{figure*}

\subsection{Alignments and Anti-alignments}
In this section, we qualitatively study what controls the final configuration of black hole-disk systems, i.e., 
alignments or anti-alignments, inspired by the arguments of \cite{King2005}.
As shown in the previous section, the (anti-)alignment timescale is generally orders of magnitude less than the 
growth timescale for black holes. Therefore, the accretion contribution to the changes of $\Jhb$ in
Equation (\ref{equ_bh}) can be safely neglected.
By denoting $\theta_{\rm inc}$ as the angle between $\Jhb$ and $\Jdb$ and using Equation (\ref{equ_torqueal}),
we have
\begin{equation}
\frac{\rd}{\rd t}\left(\Jhb\cdot\Jdb\right)=\Jhb\cdot\frac{\rd\Jdb}{\rd t}+K_2J_{\rm d}\sin^2\theta_{\rm inc}.
\end{equation}
In the meantime, above the time derivative can be alternatively written
\begin{equation}
\frac{\rd}{\rd t}\left(\Jhb\cdot\Jdb\right)=J_{\rm h}\frac{\rd J_{\rm d}}{\rd t}\cos\theta_{\rm inc}
+J_{\rm h}J_{\rm d}\frac{\rd}{\rd t}\left(\cos\theta_{\rm inc}\right),
\end{equation}
where we apply the fact that the gravitomagnetic interaction only modifies the direction of $\Jdb$
but keeps its modulus unchanged (i.e., $\rd J_{\rm h}/\rd t=0$).
A simple algebra manipulation yields
\begin{equation}
J_{\rm h}J_{\rm d}\frac{\rd}{\rd t}\left(\cos\theta_{\rm inc}\right)=K_2J_{\rm d}\sin^2\theta_{\rm inc}+
\Jhb\cdot\frac{\rd\Jdb}{\rd t}-J_{\rm h}\frac{\rd J_{\rm d}}{\rd t}\cos\theta_{\rm inc}.
\label{equ_angle}
\end{equation}
Obviously, besides the gravitomagnetic interaction, the evolution of the inclination angle $\theta_{\rm inc}$
depends on how the angular momentum of the disk $\Jdb$ changes. 

According  to the different types of disk feeding in the outer edges as illustrated in Figure~\ref{fig_sch}, 
we distinguish the two following cases.

1) {\em Fixed flows}. If the disk is steadily fed and maintains a preferred angular momentum
distribution,  one shall expect $\Jdb$, and therefore $J_{\rm d}$, to remain somewhat unchanged with time.
As a result, Equation (\ref{equ_angle}) is simplified into
\begin{equation}
\frac{\rd}{\rd t}\left(\cos\theta_{\rm inc}\right)=\frac{K_2}{J_{\rm h}}\sin^2\theta_{\rm inc}\gtrsim0.
\end{equation}
This means that, as found by previous intensive studies (\citealt{Scheuer1996, Martin2007, Perego2009}), 
a continuously fed disk always drives the hole to align with it regardless of the initial inclination.

2) {\em Free flows}. For disks without replenishment, the total angular momentum of
the system ($\Jhb+\Jdb$) is conserved. With the help of Equation (\ref{equ_torqueal}), 
it is easy to derive the following equation (see also \citealt{King2005})
\begin{equation}
\frac{\rd}{\rd t}(\cos\theta_{\rm inc})=\frac{K_2\sin^2\theta_{\rm inc}}{J_{\rm h}}
\left(1+\frac{J_{\rm h}}{J_{\rm d}}\cos\theta_{\rm inc}\right).
\label{equ_angle_free}
\end{equation}
Note that here the coefficient $K_2$ is time-dependent. Because $K_2$ is positive, once 
initially $\cos\theta_{\rm inc}\gtrsim0$, i.e., $\theta_{\rm inc}\lesssim\pi/2$, the right hand
side in the above equation is never smaller than zero. Therefore,  $\theta_{\rm inc}$ will 
decrease  continuously and the system ends up aligned. For $\theta_{\rm inc}\gtrsim\pi/2$, 
there exist two subcases:
\begin{itemize}
 \item if $J_{\rm d}/J_{\rm h}>-\cos\theta_{\rm inc}$, the inclination $\theta_{\rm inc}$
decreases;
 \item while on the contrary if $J_{\rm d}/J_{\rm h}<-\cos\theta_{\rm inc}$, 
the inclination $\theta_{\rm inc}$ increases toward $\pi$. 
\end{itemize}
In real situations, $J_{\rm d}$ generally decreases with time due to the internal 
viscous dissipation, and subsequently, the ratio
$J_{\rm d}/J_{\rm h}$ can transit from the former subcase to the latter. 
Therefore, the above conditions are indeed instantaneous for the behavior of the inclination.
We next dig into exploring the detailed
alignment/anti-alignment processes with help from numerical calculations.

\section{Numerical Scheme}
Numerically solving Equations (\ref{equ_disk}) and (\ref{equ_bh}) gives the self-consistent evolution
of the black hole accretion disk system. For this purpose, we implement a differencing 
scheme following \cite{Pringle1992}, but with some different treatment tricks.
A detailed description of our differencing scheme and robust tests of the code are given in Appendices A
and B, respectively.

In order to produce the free-torque conditions in the inner boundary, we enforce $\Lb(R_{\rm in})=0$
such that the mass and angular momentum that reaches the inner boundary are removed,
resulting in accretion onto the hole (see also \citealt{Bregman2012, Nixon2012}). 
Accordingly, the initial surface density is setup by adding an extra factor to 
allow for the torque-free condition (\citealt{Frank1992}), namely, 
\begin{equation}
\Sigma(R)=\Sigma_0 R^{-p}\left(1-\sqrt{\frac{R_{\rm in}}{R}}\right),~ 
{\rm for}~R_{\rm in}\leqslant R\leqslant R_{\rm out},
\label{equ_sur}
\end{equation}
where $R_{\rm in}$ is the inner edge of the disk.
In consideration of the inner portion of the disk aligned/anti-aligned with the hole, we set 
$\partial \elb/\partial R=0$ at $R_{\rm in}$. 

Given an initial inclination angle of $\theta_{\rm inc, 0}$, we generate the initial distribution of the direction of 
disk rings as
\begin{equation}
\elb(R)=(\sin\theta_{\rm inc, 0}, 0, \cos\theta_{\rm inc, 0}),~ 
{\rm for}~R_{\rm in}\leqslant R\leqslant R_{\rm out}.
\end{equation}
For this configuration, we presume that initially the black hole spin is oriented toward the $z-$axis. 

The outer boundary conditions at $R_{\rm out}$ are treated differently for fixed flows and free flows.
\begin{itemize}
 \item {\em Fixed flows}. Gas supply at the outer edge inherits a specified direction of angular momentum,
 that is to say $\elb(R_{\rm out})$ keeps unchanged throughout.
 The surface density at the outer boundary is fixed to $\Sigma(R_{\rm out})=\Sigma_0R_{\rm out}^{-p}
 \left(1-\sqrt{R_{\rm in}/R_{\rm out}}\right)$.
 This yields a mass supply through the outer boundary at a rate given by Equation~(\ref{equ_mdot}).
 \item {\em Free flows}. Angular momentum freely diffuses outward through $R_{\rm out}$. 
 To ensure a correct treatment of the diffusion at $R_{\rm out}$, we choose 
 the outer grid  boundary well beyond the outer disk edge (e.g., $\sim100R_{\rm out}$).
\end{itemize}

In our calculations, without specified otherwise, we adopt the fiducial values 
$\alpha=0.1$, $p=1/2$, $h=10^{-2}$, 
$R_{\rm out}=10^4R_{\rm g}$, $R_{\rm in}=6R_{\rm g}$,
and the initial black hole mass $M_\bullet=10^8M_\odot$ and spin $a=0.998$.
In principle,  $R_{\rm in}$ is determined by the marginal stable orbit of the central black hole.
However, since the inner portion of the disk is always aligned/anti-aligned with the hole,
the location of $R_{\rm in}$ does not affect the gravitomagnetic torque and hence the alignment rate.
Moreover, a relatively larger $R_{\rm in}$ helps to improve the time consumption of the computations in view of the 
timestep size limited by viscous timescales (see Equation (\ref{equ_tw})).
To assess the processes of alignments and anti-alignments, we adjust the two free parameters:
the mass fraction of the disk to the black hole $f_m$,
which determines the surface density of the disk in Equation (\ref{equ_sur}) and hence
the mass accretion; and the initial inclination angle between the disk and the hole $\theta_{\rm inc, 0}$.
We setup the grid with 202 points and evolve the equation set for a time of $1.5t_\nu$, where
$t_\nu=2\times10^6$yr is the viscous time at $R_{\rm out}$ (see Appendix A for details).

\section{Results}
\subsection{Evolution of Warped Accretion Disks}
\begin{figure}[t!]
\centering
\includegraphics[angle=-90.0, width=0.38\textwidth]{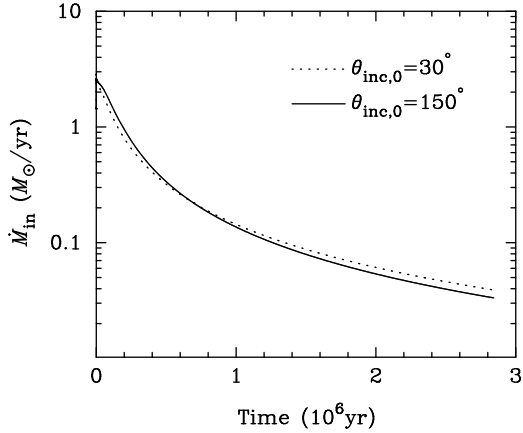}
\caption{Mass accretion rate onto black holes  ($\dot M_{\rm in}$) with time for free flows.}
\label{fig_mdot1}
\end{figure}

\begin{figure}[t!]
\centering
\includegraphics[angle=-90.0, width=0.38\textwidth]{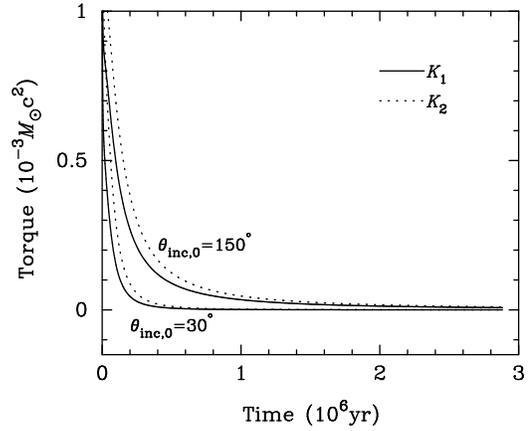}
\caption{Components of the gravitomagnetic torque in Equation (\ref{equ_torqueal}) for free flows. 
$K_1$ corresponds to the precessions and $K_2$ corresponds to aligning the system. 
Note that both $K_1$ and $K_2$ are positive.}
\label{fig_torque}
\end{figure}

 We define the inclination of disks at radius $R$ with respect to black holes  as 
\begin{equation}
\theta_{\rm inc}(R)=\cos^{-1}\frac{\Jhb\cdot\Lb(R)}{|\Jhb||\Lb(R)|}.
\end{equation}
In Figure~\ref{fig_profile1}, we illustrate how the shape of warped accretion disks evolves with times
for free flows. The initial inclination angles are $\theta_{\rm inc, 0}=30^\circ$ in the upper
panel and $\theta_{\rm inc, 0}=150^\circ$ in the bottom panel, respectively.
We can find that an inner flat disk rapidly forms for all cases and is aligned to the black hole for 
$\theta_{\rm inc, 0}=30^\circ$ and anti-aligned for $\theta_{\rm inc, 0}=150^\circ$.
The size of the inner flat disk continuously grows until roughly $10^2R_{\rm g}$, corresponding to
the warp radius $R_{\rm w}$ defined in Equation (\ref{equ_rwarp}).
Warp propagation reaches the outer edge of the disk within a timescale of $\sim0.01t_\nu$, consistent with
the estimate from Equation (\ref{equ_tw2}). 
The disk for $\theta_{\rm inc, 0}=30^\circ$ gradually approaches
full alignments; however, the alignment rate is prone to  severe attenuation with time. 
The disk inclination is almost unchanged from $1.0$ to $1.5t_\nu$.  
At the end of the numerical calculation $1.5t_\nu$,
the disk maintains an inclination of $\theta_{\rm inc}\sim2^\circ$ at the outer edge.
Henceforth, it hardly aligns with the black hole fully. This is because of a severe depletion of the 
surface density due to mass accretion (see Equation (\ref{equ_td})). 
In Figure~\ref{fig_mdot1}, it is verified that the mass accretion onto the black hole
decreases rapidly with time  by more than one order of magnitude as a result of disk depletion.
This implies that if there is no gas supply, the luminosity of the accretion disk will fade out less than $10^6$ yr.
We further calculate the components of the gravitomagnetic torque that the whole disk exerts on the hole 
$K_1$ and $K_2$ as defined in Equation (\ref{equ_torqueal}). 
As shown in Figure \ref{fig_torque}, it is confirmed that irrespective of the initial inclinations, both $K_1$ and $K_2$ are positive. 
Moreover, $K_1$ and $K_2$ are in the same order of magnitude (\citealt{Scheuer1996, King2005, Martin2007}).

The system for $\theta_{\rm inc, 0}=150^\circ$ evolves toward anti-alignment in a similar fashion,
as expected since initially $J_{\rm d}/J_{\rm h}=0.78<\cos\theta_{\rm inc,0}=0.87$.
However, the inclination $\theta_{\rm inc}$ at $R_{\rm out}$ has changed only from $150^\circ$
to $\sim160^\circ$ at the end of the calculation, which is very inefficient compared to the case of
$\theta_{\rm inc, 0}=30^\circ$. The corresponding reason is that, as shown 
in Equation (\ref{equ_angle_free}), the change of $\theta_{\rm inc}$ from the 
gravitomagnetic torque is proportional to $\propto(1+J_{\rm h}/J_{\rm d}\cos\theta_{\rm inc})$. 
Its value for $\theta_{\rm inc, 0}=150^\circ$ is smaller by an order of magnitude than
that for $\theta_{\rm inc, 0}=30^\circ$.

\begin{figure*}[t!]
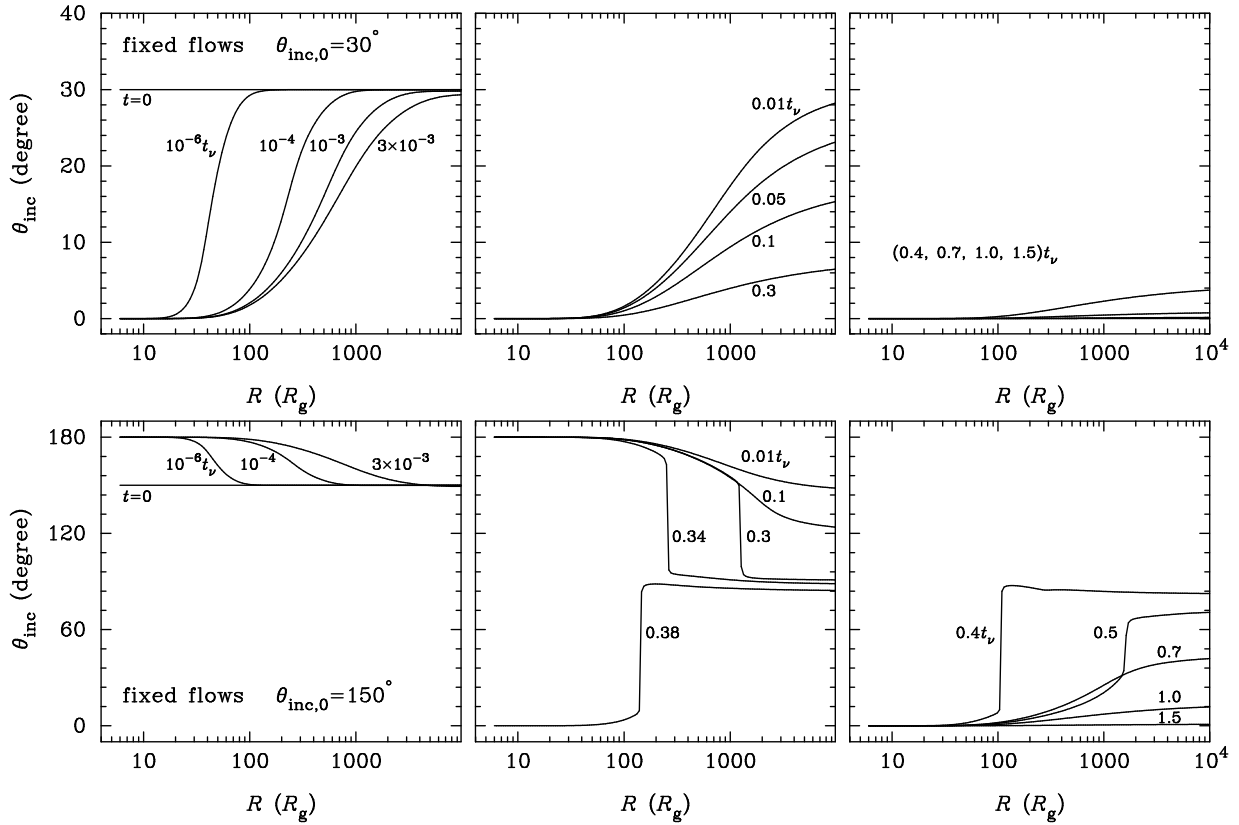

 \centering
  \includegraphics[angle=-90.0, width=0.9\textwidth]{profile12.ps}
 \includegraphics[angle=-90.0, width=0.9\textwidth]{profile22.ps}
 \caption{Same as Figure \ref{fig_profile1} but for fixed flows.}
  \label{fig_profile2}
\end{figure*}

The cases for fixed flows are illustrated in Figure~\ref{fig_profile2}.
By analogy to free flows, warps spread to the outer boundary within $0.01t_\nu$; 
however, the subsequent evolution is somewhat different. The system 
with $\theta_{\rm inc, 0}=30^\circ$ quickly achieves full alignment after a time less than $1.0t_\nu$.
For $\theta_{\rm inc, 0}=150^\circ$, the disk shape shows complicated behaviors.
The inner anti-aligned angular momentum is transported outward by warp propagation and 
confronts the  angular momentum carried by the feeding matter.
As a result, an abrupt discontinuity of the inclination appears and 
the disk breaks into two parts: an inner nearly anti-aligned portion and 
an outer misaligned portion (see also \citealt{Lodato2006, Nixon2012, Nixon_etal2012}). 
The location of the discontinuity transfers inward with time and the inner portion will be eventually 
swallowed by the hole due to mass accretion. In the meantime, the black hole is driven to align with 
the outer portion progressively. Around the time $0.34-0.38t_\nu$, after the inner anti-aligned portion of the disk
is consumed, the newly formed inner disk turns to be aligned with the black hole. At the end of the calculation
$1.5t_\nu$, there is complete alignment between the disk and the black hole.

\begin{figure}[t!]
\centering
\includegraphics[angle=-90.0, width=0.38\textwidth]{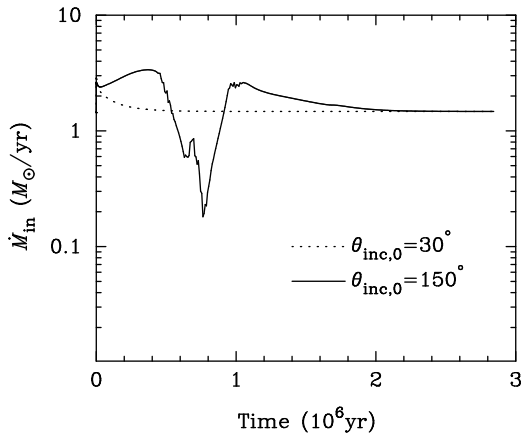}
\caption{Mass accretion rate onto black holes ($\dot M_{\rm in}$) with time for fixed flows.
The tiny ripples in the $\theta_{\rm inc,0}=150^\circ$ line is due to the numerical effect.}
\label{fig_mdot2}
\end{figure}

The mass accretion rates onto the black holes for fixed flows are plotted in Figure~\ref{fig_mdot2}.
From Equation (\ref{equ_vr}), the role of warping increases the inflow velocity and therefore 
enhances the mass accretion rate. As expected, for $\theta_{\rm inc, 0}=30^\circ$, 
the accretion rate onto the hole ($\dot M_{\rm in}$) has a mildly decreasing trend by a factor of two
at the beginning in response to the approaching alignment of the system. Then it 
maintains a steady value of $\sim 1M_\odot{\rm yr}^{-1}$, equal to the gas supply rate at the outer 
boundary. Quite differently, for $\theta_{\rm inc, 0}=150^\circ$ there exist two peaks of the mass
accretion rate at times $\sim0.2t_\nu$ and $\sim0.5t_\nu$ due to the large magnitude of warping developed
at these moments (see the bottom panel of Figure~\ref{fig_profile2}). While the inner anti-aligned disk portion
is being consumed, the accretion rate becomes significantly attenuated by orders of magnitude and reaches
the minimum ($\dot M_{\rm in}\rightarrow0$) around $0.34-0.38t_\nu$. 
Nevertheless, the mass accretion onto the hole eventually approaches a steady rate, the same as for 
$\theta_{\rm inc, 0}=30^\circ$. It is worth pointing out that the black hole undergoes retrograde accretion
for a time $\sim0.35t_\nu$, after which it transits to prograde accretion.


\begin{figure*}[t!]
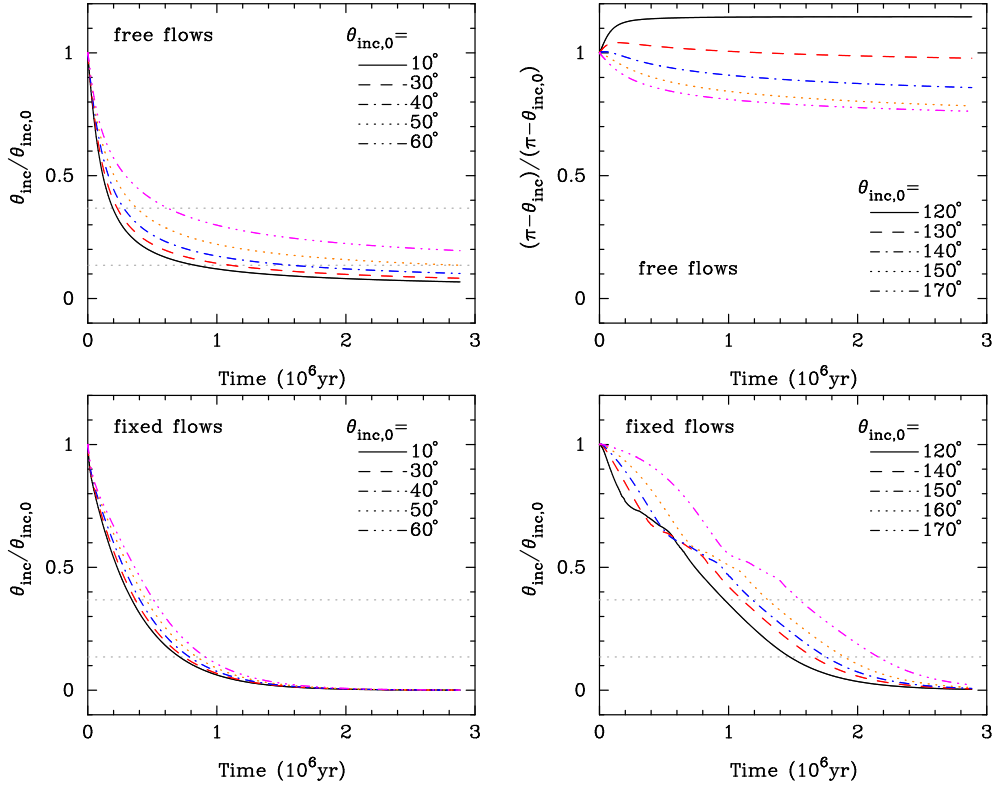

\centering
\includegraphics[angle=-90.0, width=0.35\textwidth]{timec11.ps}~~~~~~
\includegraphics[angle=-90.0, width=0.35\textwidth]{timec12.ps}
\includegraphics[angle=-90.0, width=0.35\textwidth]{timec21.ps}~~~~~~
\includegraphics[angle=-90.0, width=0.35\textwidth]{timec22.ps}
\caption{Evolution of the inclination between the holes and the disks for two classes of
boundary conditions: (upper) free flows and (bottom) fixed flows.
Two horizontal dashed lines correspond to $\theta_{\rm inc}/\theta_{\rm inc,0}=e^{-1}$ and $e^{-2}$,
respectively.}
\label{fig_time}
\end{figure*}
\begin{figure*}[t!]
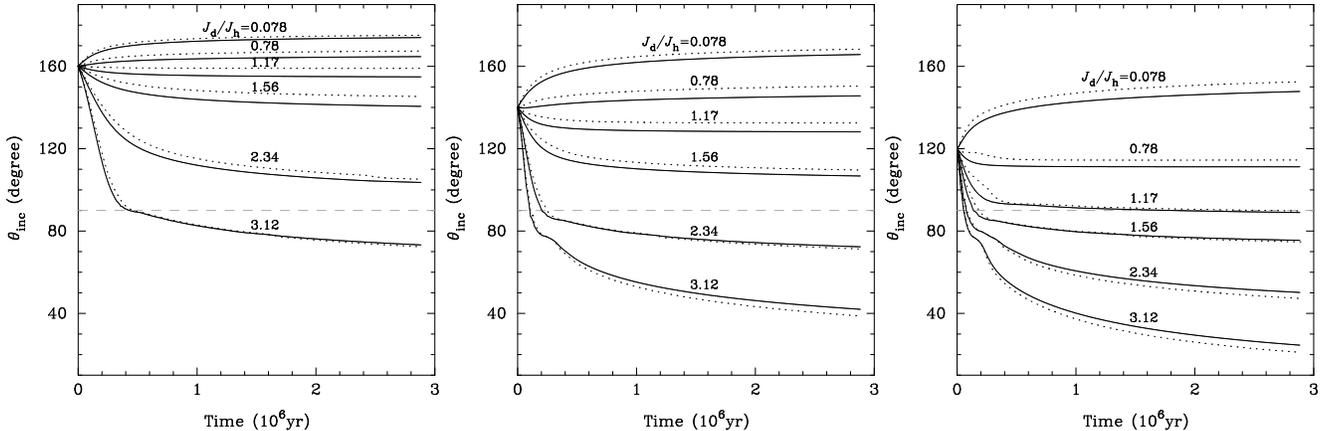

\centering
\includegraphics[angle=-90.0, width=0.32\textwidth]{cond2.ps}
\includegraphics[angle=-90.0, width=0.32\textwidth]{cond1.ps}
\includegraphics[angle=-90.0, width=0.32\textwidth]{cond3.ps}
\caption{Evolution of the inclination angle between the angular momenta of the black holes and the disks 
for free flows with different initial $J_{\rm d}/J_{\rm h}$, which is
determined by the parameter $f_m$ as in Equation (\ref{equ_ratio}). 
The initial inclination angles are $\theta_{\rm inc,0}=160^\circ$ (top), $140^\circ$ (middle), and $120^\circ$ (bottom).
Dotted lines correspond to the angles in terms of the angular momentum of the disk just within $R_{\rm out}$
(excluding the parts that outflow through $R_{\rm out}$).
Gray horizontal dashed lines correspond to $\theta_{\rm inc}=90^\circ$. }
\label{fig_anti}
\end{figure*}
\subsection{Alignment and Anti-alignment Timescales}
Figure~\ref{fig_time} plots the evolution of the inclination angles between the black hole and the disk 
for two classes of accretion flows. Here the inclination angle is defined by
\begin{equation}
\theta_{\rm inc}=\cos^{-1}\frac{\Jdb\cdot\Jhb}{|\Jdb||\Jhb|},
\label{equ_inc}
\end{equation}
where $\Jdb$ is the total angular momentum of the disk.
The upper left panel shows $\theta_{\rm inc}/\theta_{\rm inc,0}$
against time with $\theta_{\rm inc,0}=(10^\circ, 30^\circ, 40^\circ, 50^\circ, 60^\circ)$ from bottom to top
for free flows.
The timescales that the inclination angle decreases to $1/e$ of its initial value generally lie
at several $10^{5}$yr, consistent with the magnitude estimate in Equation (\ref{equ_tal}).
However, the decaying rate of the inclination angle rapidly slows down with time because of the depletion of the disk
(almost all of the disk angular momentum is transported to the large radius).
After $\sim10^6$ yr, the inclinations for all  the cases are prone to be frozen, indicating that
the disk shall hardly approach complete alignment. This further confirms the results in the previous section. 
The upper right panel plots $(\pi-\theta_{\rm inc})/(\pi-\theta_{\rm inc,0})$ with  $\theta_{\rm inc,0}=
(120^\circ, 130^\circ, 140^\circ, 150^\circ, 170^\circ)$ 
from top to bottom for free flows. Overall, although quite inefficient, the inclination angles have a tendency to approach
$\theta_{\rm inc}\rightarrow180^\circ$ except for $\theta_{\rm inc, 0}=120^\circ$, 
implying a trend of anti-alignment between the hole and the disk. The system for 
$\theta_{\rm inc, 0}=120^\circ$ tends toward alignment instead of anti-alignment because
$J_{\rm d}/J_{\rm h}>-\cos\theta_{\rm inc}$ is always retained throughout the calculation (see Section 3.3).
Interestingly, there is an increasing trend at the beginning for $\theta_{\rm inc, 0}=130^\circ$ and 
$140^\circ$, because $J_{\rm d}/J_{\rm h}>-\cos\theta_{\rm inc}$ initially, which, however,  transits to
$J_{\rm d}/J_{\rm h}<-\cos\theta_{\rm inc}$ later on.
Again, the alignment rates are significantly slower compared with the counterparts in the upper left panel.

The bottom panels of Figure~\ref{fig_time} show the results for fixed flows. 
The timescales where $\theta_{\rm inc}$ decreases to $1/e$ of its initial value  are 
also generally several $10^{5}$yr for $\theta_{\rm inc,0}<90^\circ$ (bottom left panel) and almost insensitive to
$\theta_{\rm inc,0}$. Within a time of $1.0t_\nu$, the entire disks have been in complete alignment with
the holes.
In the bottom right panel, $\theta_{\rm inc}$ decreases with time even though $\theta_{\rm inc,0}>90^\circ$
as  was expected through the previous theoretical analysis.
There are inflection points in the evolution curves of $\theta_{\rm inc}$ that correspond to when
the disks break into two parts as illustrated in Figure~\ref{fig_profile2}.
Generally, the timescales of the decaying rate for $\theta_{\rm inc,0}>90^\circ$ (bottom right panel) are relatively longer 
compared to these for $\theta_{\rm inc,0}<90^\circ$  (bottom left panel) due to the same reasons for free flows. However,
because the disks are replenished and therefore the gravitomagnetic torques from the disks are not attenuated, 
the decreasing rates of $\theta_{\rm inc}$ are clearly more rapid than their counterparts 
(top right panel) for free flows. 

In summary, for fixed flows in which disks are continually fed, alignments always occur regardless
of the initial inclination.  The alignment timescale is of an order of $2\times10^6$ yr. For free flows in which there is
no gas supply, both alignments and anti-alignments could occur. However, as a result of the notable depletion of 
the disk's surface density at a timescale of $10^6$yr, the alignment/anti-alignment
rate is significantly reduced, leading the disks to be not completely aligned/anti-aligned.

\subsection{Conditions for Anti-alignments}
Previous sections show that anti-alignment between black holes and disks is possible only when
the initial inclination angle is $\theta_{\rm inc, 0}>90^\circ$ for free flows.
To assess how the final configuration of the system depends on the free parameters,  
in this section we evolve the equation set given different parameters $f_m$ and $\theta_{\rm inc,0}$
only for free flows.
We note that the parameter $f_m$ determines the initial total angular momentum of the disks as in
Equation (\ref{equ_ratio}).
Figure~\ref{fig_anti} plots the evolution of inclination angle $\theta_{\rm inc}$
between $\Jhb$ and $\Jdb$ with $f_m/10^{-2}=$0.1, 1, 1.5, 2, 3, 4, and 6, corresponding to the initial ratios 
$J_{\rm d}/J_{\rm h}=0.078$, 0.78, 1.17, 1.56, 2.34, and 3.12, respectively. 
The initial inclination angles are $\theta_{\rm inc,0}=160^\circ$, $140^\circ$, and
$120^\circ$ from top to  the bottom panel. Since gas diffuses outward 
through $R_{\rm out}$ for free flows, there is an outflow of angular momentum at $R_{\rm out}$ accordingly.
Here $\Jdb$ includes these aspects of angular momentum.
For comparison purposes, we also calculate the inclination angles in terms 
of the angular momentum of the disk just within $R_{\rm out}$, plotted by dotted lines in Figure~\ref{fig_anti}. 
As can be seen, the evolution trends of $\theta_{\rm inc}$ depend both on the ratios 
$J_{\rm d}/J_{\rm h}$ and the initial inclinations.
Specifically, there are three prominent features as follows. 

1) Consistent with the prediction in Section 3.3,
the inclination angles $\theta_{\rm inc}$ generally
go up when the initial ratios $J_{\rm d}/J_{\rm h}<-\cos\theta_{\rm inc, 0}$, but otherwise,
drop off. We note that in the middle panel of Figure~\ref{fig_anti} 
the inclination of the system with an initial ratio $J_{\rm d}/J_{\rm h}=0.78$, larger than $-\cos\theta_{\rm inc, 0}=0.76$,
descends mildly  at the very beginning and then increases toward $180^\circ$. This is
because the ratio $J_{\rm d}/J_{\rm h}$ rapidly decreases to $J_{\rm d}/J_{\rm h}<-\cos\theta_{\rm inc}$.

2) The inclination for $\theta_{\rm inc, 0}=160^\circ$ begins to enter $\theta_{\rm inc}<90^\circ$ when
the initial ratio $J_{\rm d}/J_{\rm h}$ is larger than $\sim3$. 
This means that the system will approach alignment, and most importantly, the black holes undergo a transition
from retrograde  to prograde accretion.
Similarly, the critical ratios for 
$\theta_{\rm inc, 0}=140^\circ$ and $120^\circ$ are $\sim2.0$ and $1.2$, respectively.

3) Without replenishment, the disks are greatly depleted on a timescale of $10^6$ yr.  
As a result, the gravitomagnetic torques are reduced with time and 
the inclination angles are almost frozen to values not equal to $0^\circ$ or $180^\circ$ at the end of calculations. 
In other words, the systems hardly achieve full alignment or anti-alignment.

On the basis of the angular momentum conservation of the system (free flows in the present study), 
\cite{King2005} proposed
that anti-alignments will occur provided that the initial angle between the angular momenta
of the hole and the disk satisfy $J_{\rm d}/J_{\rm h}\lesssim-2\cos\theta_{\rm inc, 0}$. 
Applying this formula to the present cases, for $\theta_{\rm inc, 0}=160^\circ$, anti-alignment 
occurs when the ratio obeys $J_{\rm d}/J_{\rm h}\lesssim1.88$ initially; for 
$\theta_{\rm inc, 0}=140^\circ$ and $120^\circ$, the corresponding critical ratios 
are $1.53$ and $1.0$, respectively. Our results in Figure~\ref{fig_anti} are in considerable
disagreement with these values. We ascribe such discrepancy to the neglect of  the disk's depletion
in the previous work. 
The ratio between the timescales for 
(anti-)alignments and the disk's depletion
is of an order of unity for fiducial parameters from Figure~\ref{fig_time}.
During accretion onto the black hole, the disk's mass is gradually consumed
and the majority of the disk's angular momentum is transported to the large radius
(\citealt{Frank1992}). 
Note that the gravitomagnetic torque is in proportion to $R^{-3}$,
therefore the (anti-)alignment rate will be greatly reduced. This effect leads to some systems with
ratios of $J_{\rm d}/J_{\rm h}>-2\cos\theta_{\rm inc, 0}$ still maintaining $\theta_{\rm inc}>90^\circ$
at the end of the calculations in Figure~\ref{fig_anti}.

\section{Discussions}
\subsection{Implications for Black Hole Spin}
In addition to fueling black holes, accretion also adds angular momentum carried by matter
to black holes and accordingly modifies their spin. How spin changes depends on the orbits
at the inner edge of the accretion disks. Specifically, if the disk is aligned with the black hole, 
prograde accretion will spin up the hole; otherwise, retrograde accretion spin down the hole.
The black hole will become maximally rotating ($a\rightarrow1$) when its mass is doubled provided prograde 
or alternative retrograde accretion is preferably retained for a sufficient amount of time (\citealt{Thorne1974}).
In view of the episodic and random activities of black hole accretion (\citealt{Wang2006, Li2010}), 
it is crucial to model the spin evolution whether the accretion in each episode proceeds progradely or retrogradely.

Our results obtained above imply that the configurations of the accretion disks depend
on feeding them at the outer boundaries. If the disks are continually 
fed for a long enough lifetime (say, e.g., $10^7$yr), 
alignments always occur. In this case, prograde accretion is the dominated fashion.
We expect that after a series of activities, black holes will be close to maximal rotation.
On the other hand, if there is no gas feeding,  both alignments and anti-alignments
are possible, depending on the initial inclinations and the ratios $J_{\rm d}/J_{\rm h}$
(see Section 5.3). One has to follow the probabilities of alignments and anti-alignments over 
episodes using the method presented here to determine spin evolution.
An incorporation of the present calculations into semi-analytical models of galaxy formation
and evolution will shed light on the existing studies in the field 
(e.g., \citealt{Lagos2009, Fanidakis2011, Barausse2012,Volonteri2012, Dotti2013}),
and provide more realistic information on SMBH spin. 

Note that black hole accretion only lasts a lifetime of $\sim10^6$yr as a result of 
disk depletion without feeding.\footnote{The lifetime can be generally estimated as follows: 
for black holes accreting at the Eddington limit $\dot M_{\rm Edd}\sim 2
M_8M_\odot{\rm yr}^{-1}$, where $M_8=M_\bullet/10^8M_\odot$, the accretion disks
with masses of $\sim (H/R)M_\bullet=10^6M_8M_\odot$ will be consumed at
$\sim10^6$yr, where $H/R\approx10^{-2}$.}  
Indeed, the lifetime of feeding disks directly determines the 
observable episodic lifetimes of AGNs.
In this sense, observational measurements
of the episodic lifetimes of AGNs can place useful constraints on the SMBH spin evolution.
Unfortunately, thus far the episodic
lifetime of AGNs remains elusive though some endeavor has been put forth
(see a review of \citealt{Martini2004}). The most compelling method---proximity effects estimated the 
episodic lifetimes in a broad range of $\sim10^6-10^8$yr, varying
from study to study (\citealt{Kirkman2008, Goncalves2008, Furlanetto2011}).
Future observations from large sample quasar surveys with a variety of techniques 
would offer promising measurements on the episodic lifetimes and help to
understand the cosmological evolution of SMBH spin.

On the other hand, we turn over the issue so as to constrain the episodic lifetimes of AGNs
from the otherwise obtained SMBH spin evolution. Recently, quantifying the radiative 
efficiency of mass accretion through SMBH demography strongly suggests that 
SMBHs are spinning down with cosmic time since $z\sim2$ (\citealt{Wang2009, Li2011, Li2012};
see also the discussion of \citealt{Zhang2012}).
Combined with the present results, this indicates that the episodic lifetime of AGNs
must not exceed several $10^6$ yr and probably have cosmological evolution as well.
More detailed modelings on the connections between SMBH spin evolution and the episodic lifetime
of AGNs are highly deserved in future works.

It is worth stressing that we presume black holes to be accreting at 
the Eddington limit in the calculations. The timescale for (anti-)alignments between the disks 
and the holes is in proportion to the reciprocal of the mass accretion rate (see Equation (\ref{equ_tal})).
Therefore, if the presumed mass accretion rate is relaxed, the above proposed limit on the AGN lifetime
changes in the same manner accordingly.

\subsection{Uncertainties}
In our calculations, the inner edge of the disk is fixed at $R_{\rm in}=6R_{\rm g}$.
Considering that the inner portion of the disk within $R_{\rm w}$ always remains
aligned with the hole, this does not affect the gravitomagnetic torque on the hole and therefore
the alignment rate. On the other hand, during the alignment course, the accreted mass
and angular momentum are negligible with those of the black hole. Hence, 
the results presented in this paper are insensitive to the location of the inner boundary.
For the sake of simplicity, the disk's outer edge (limited by the self-gravitating) 
is fixed at $R_{\rm out}=10^4R_{\rm g}$.
In reality, the self-gravitating radius depends on the accretion rate, but 
generally lies at a range of $\sim10^3-10^4R_{\rm g}$ for standard accretion disks 
(e.g., \citealt{Goodman2003, King2008}). The location of $R_{\rm out}$ just 
determines the ratio of $J_{\rm d}/J_{\rm h}\propto R_{\rm out}^{1/2}$ as in Equation (\ref{equ_ratio}), 
but does not affect the gravitomagnetic torque and (anti-)alignment timescale
as in Equations (\ref{equ_torque}) and (\ref{equ_tal}). The overall results remain unchanged.

The viscosities for accretion disks are treated using the first-order approximations given by
\cite{Ogilvie1999}. The nonlinear fluid effects that appear for large amplitude warp would
modify the expressions of the effective viscosities (\citealt{Ogilvie1999, Lodato2010}). 
We expect this to become significant
when the disks break as shown in the bottom panels of Figure~\ref{fig_profile2}. 
In this case, however, it is unclear whether the evolution equation set adopted here remains adequate.
Dedicated investigations, including nonlinear fluid dynamics, are beyond the scope
of the present study. We defer this to future  sophisticated numerical simulations.

Finally, we consider the accretion disks under the Newtonian timespace and treat the gravitomagnetic
interaction in the weak-field limit (the Lense-Thirring precession). The general relativistic calculations
are of somewhat less importance because the warp radius $R_{\rm w}$ is far beyond the gravitational radius 
$R_{\rm g}$ (see Equation (\ref{equ_rwarp})).
\section{Conclusions}
We numerically solve the evolution equation set of spinning black holes and their warped accretion disks.
To mimic the realistic accretion processes, we consider two classes of accretion disks in terms of
the outer boundary conditions: {\em fixed flows} for which the disk is continually fed with 
a preferred angular momentum through its outer edge; and {\em free flows} for which
there is no gas supply and the disk diffuses freely at its outer edge (see Figure~\ref{fig_sch}).
Our main results are as follows.
\begin{enumerate}
 \item For disks continually fed,  alignments between the holes and disks always occur
 at a time of several $10^5$ yr regardless of the initial inclinations. 
 If the initial inclination angles are $\theta_{\rm inc, 0}>\pi/2$, black hole accretion transits from retrograde fashion
  to prograde fashion after a time of several alignment timescales. The mass growths during these two phases
  are comparable.
 
 \item For disks without a gas supply, both alignments and anti-alignments are possible, depending on
 the initial inclinations $\theta_{\rm inc,0}$ and the ratios $J_{\rm d}/J_{\rm h}$. 
 It is worthwhile to point out that the lifetime of black hole accretion at the Eddington limit
 in such cases last $\sim10^6$ yr
 due to the depletion of the accretion disks. This has important influences on the (anti-)alignment processes.
 
 \item Black hole spin is closely linked to the manners of mass accretion, namely, progradely or
 retrogradely. Careful inclusion of disk feeding at the outer boundaries is highly 
 important to modeling spin evolution in the context of episodic and random activities of SMBHs. 
 The episodic lifetime of AGNs, which can be measured with a variety of techniques, 
 is directly determined by the lifetime of disk feeding. We therefore propose that
 there would be a close connection between the black hole spin and the episodic lifetime
 of AGNs, which deserves a comprehensive investigation in future works.
\end{enumerate}

\acknowledgements{
This research is supported by NSFC-11133006, 11173023, and 11233003, and a 973 project (2009CB824800). 
The numerical calculations in this work used the computer clusters at the 
Institute of High Energy Physics.}

\appendix
\section{A. Differencing The Equation Set}
We construct the differencing scheme of Equations (\ref{equ_disk}) and (\ref{equ_bh}) 
generally following \cite{Pringle1992},
but with some modifications. For the sake of completeness, we summarize our implementation here.

The spatial domain is discretized into ($I+2$) logarithmic grid points: 
$R_i=R_{\rm in}e^{i\Delta x}$, for $i=0,..., I+1$, with $\Delta x$ the equal spacing of the logarithmic grid.
The points $R_0$ and $R_{\rm I+1}$ represent the inner and outer grid boundaries, respectively.
If we define the advective velocity 
\begin{equation}
V_{\rm adv}=\frac{3}{2}\frac{\nu_1}{R}-\nu_2 R \left|\frac{\partial \elb}{\partial R}\right|^2,
\end{equation}
and use Equation (\ref{equ_vr}),  Equation (\ref{equ_disk}) can be written as
\begin{equation}
\frac{\partial \Lb}{\partial t}=-\frac{1}{R}\frac{\partial}{\partial R}\left(V_{\rm adv}R\Lb\right)
+\frac{1}{R}\frac{\partial}{\partial R}\left[3R\frac{\partial}{\partial R}(\nu_1 L)\boldsymbol{\ell}
+\frac{1}{2}\nu_2 R L\frac{\partial \boldsymbol{\ell}}{\partial R}\right]
+\frac{1}{R}\frac{\partial}{\partial R}\left[\nu_3 R\Lb\times\frac{\partial \boldsymbol{\ell}}{\partial R}\right]
+\frac{2G}{c^2}\frac{\Jhb\times\Lb}{R^3}.
\label{equ_disk_app}
\end{equation}
Notationally, we use a superscript $n$ to denote the timestep and a subscript $i$ to denote the spatial
grid point. The differencing scheme of Equation (\ref{equ_disk_app}) is then built as follows 
(see also \citealt{Bregman2012})
\begin{eqnarray}\nonumber
\frac{\Lb_{i}^{n+1}-\Lb_{i}^{n}}{\Delta t}&=&
-\frac{(V_{\rm adv})_{k+1}^nR_{k+1}\Lb_{k+1}^n - (V_{\rm adv})_{k}^n R_{k}\Lb_{k}^n}{R_i^2\Delta x}
+\frac{3\left\{\left[(\nu_1 L)_{i+1}^n-(\nu_1 L)_i^n\right]\elb_{i+1/2}^n -
\left[(\nu_1 L)_{i}^n-(\nu_1 L)_{i-1}^n\right]\elb_{i-1/2}^n\right\}}{R_i^2(\Delta x)^2}\\\nonumber
&&+\frac{(\nu_2 L)_{i+1/2}^{n}(\elb_{i+1}^n-\boldsymbol{\ell}_{i}^n)
-(\nu_2 L)_{i-1/2}^{n}(\elb_{i}^n-\elb_{i-1}^n)}{2R_i^2(\Delta x)^2}
+\frac{(\nu_3 \Lb)_{i+1/2}^n\times(\elb_{i+1}^n-\elb_{i}^n)
-(\nu_3 \Lb)_{i-1/2}^n\times(\elb_{i}^n-\elb_{i-1}^n)
}{R_i^2(\Delta x)^2}\\
&& + \frac{2G}{c^2}\frac{\mathbf{J}_{\rm h}^n\times\Lb_i^n}{R_i^3},
\label{equ_disk_df}
\end{eqnarray} 
where $\Delta t$ is the timestep size, and the advective term is treated using upstream differencing: 
for $V_{\rm adv}>0$, $k=i-1$; whereas for $V_{\rm adv}<0$, $k=i$. The advective velocity is calculated by
\begin{equation}
(V_{\rm adv})_i^n=\frac{1}{R_i}\left[\frac{3}{2}(\nu_1)_{i}^n - 
(\nu_2)_{i}^n\left|\frac{\boldsymbol{\ell}_{i+1}^n-\boldsymbol{\ell}_{i-1}^n}{2\Delta x}\right|^2\right].
\end{equation}
Similarly, the differencing scheme of Equation (\ref{equ_bh}) is built as
\begin{eqnarray}
\frac{\mathbf{J}_{\rm h}^{n+1}-\mathbf{J}_{\rm h}^{n}}{\Delta t}=(\dot M_{\rm in} \mathbf{j}_{\rm in})^n
-\frac{4\pi G}{c^2}\mathbf{J}_{\rm h}^n\times\sum_{i=1}^{I} \frac{\Lb_i^n}{R_i} \Delta x,
\end{eqnarray}
where the mass accretion rate onto the hole is calculated as $(\dot M_{\rm in})^n=-2\pi (R V_R \Sigma)_1^n$ and
the specific angular momentum at the inner edge is calculated as 
$\mathbf{j}_{\rm in}^n=(R^2\Omega)_1^n \boldsymbol{\ell}_1^n$.
As described in Section 4, at the inner boundary, we take $\Lb_0^n=0$, and $\elb_0^n=\elb_1^n$; 
at the outer boundary, for {\em free flows}, we take $\Lb_{I+1}^n=(\nu_1 \Lb)_{I}^n/(\nu_1)_I^n$;
for {\em fixed flows} we take
$L_{I+1}^n=R_{I+1}^2\Omega_{I+1}\Sigma_0 R_{I+1}^{-p}(1-\sqrt{R_{\rm in}/R_{I+1}})$
and fix $\elb_{I+1}^n=(\sin\theta_{\rm inc, 0}, 0, \cos\theta_{\rm inc, 0})$. 
This enforces a coherent mass supply at a rate given roughly by Equation (\ref{equ_mdot}).
The timestep size is adjusted for every step according to
\begin{eqnarray}
\Delta t=\frac{1}{2}\min\left\{\Delta t_1, \Delta t_2, \Delta t_3, \Delta t_{\rm CFL},
\Delta t_{\rm LT}, \Delta t_{\rm h}\right\}, 
\end{eqnarray}
where 
\begin{equation}
\Delta t_j=\min_{1<i<I}\frac{(R_i\Delta x )^2}{(\nu_j)_i}~~~{\rm for}~ j=1, 2, 3, 
\end{equation}
\begin{equation}
\Delta t_{\rm CFL}=\min_{1<i<I}\frac{R_i\Delta x}{(V_{\rm adv})_i},~~~
\Delta t_{\rm LT}=\min_{1<i<I} \frac{L_i}{|\mathbf{T}_{\rm LT}|_i}, ~~~{\rm and}~~~
\Delta t_{\rm h}=J_{\rm h}\left|\frac{4\pi G}{c^2}\mathbf{J}_{\rm h}
\times\sum_{i=1}^{I} \frac{\Lb_i}{R_i} \Delta x\right|^{-1}. 
\end{equation}
\begin{figure}[t!]
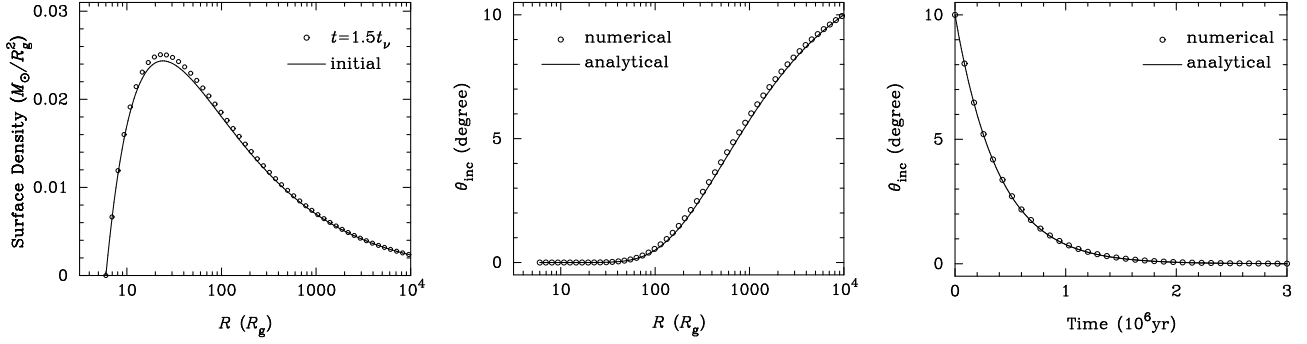

\centering
\includegraphics[angle=-90.0, width=0.31\textwidth]{test3.ps}~~~~
\includegraphics[angle=-90.0, width=0.3\textwidth]{test1.ps}~~~~
\includegraphics[angle=-90.0, width=0.3\textwidth]{test2.ps}
\caption{To verify the validity of our numerical code, three tests are performed (see the text for details).
Left: the surface density of a flat disk ($\theta_{\rm inc, 0}=0$ or $\pi$) at the beginning and end
of the calculations.
Middle: inclination profile of the disk in a steady state with the outer inclination angle fixed at 
$\theta_{\rm inc}=10^\circ$. 
Right: Evolution of inclination angle between the hole and the disk with an initial angle  of
$\theta_{\rm inc, 0}=10^\circ$.}
\label{fig_test}
\end{figure}

\section{B. Validity of the Numerical Code}
To verify the validity of our numerical code, we perform three tests. 
(1) We run the code upon a flat disk ($\theta_{\rm inc, 0}=0$ or $\pi$) for a time of $1.5t_\nu$.
The left panel of Figure~\ref{fig_test} plots the surface density of the disk at the beginning and end of
the calculations. There are tiny deviations less than 2\% due to the dissipation of the numerical scheme.
(2) For small warping, there exist analytical solutions if disks are continually fed at the outer edges, 
analogous to fixed flows defined in the present paper (\citealt{Scheuer1996, Martin2007, Chen2009}). 
We only evolve Equation (\ref{equ_disk}) with a sufficient amount of time
using the differencing scheme described above and obtain a steady shape of the disk. 
In the middle panel of Figure~\ref{fig_test}, we compare the numerically obtained inclination profile of the disk
with the analytic solution given by Equation (24) in \cite{Martin2007}. A complete match can be found.
(3) Then initializing the disk with the above shape, we go on to evolve Equations (\ref{equ_disk})
and (\ref{equ_bh}) simultaneously to obtain the time-dependent inclination angle between the hole and the disk.
The right panel of Figure~\ref{fig_test} shows that our numerical calculations are in exact 
agreement with the analytical solutions given by Equation (52) in
 \cite{Martin2007}. These three tests indicate a validity of our numerical code.

\end{document}